\documentclass[onecolumn]{emulateapj}

\slugcomment{Astro-ph archives (submitted: 7$^{th}$ May 2013)}

\shorttitle{X1.8 SOLAR ERUPTIVE FLARE AND ASSOCIATED FILAMENT ERUPTION}
\shortauthors{JOSHI ET AL.}

\begin{document}



\title{RHESSI and TRACE observations of multiple flare activity in AR 10656 and associated filament eruption} 



\author{Bhuwan Joshi, Upendra Kushwaha}
\affil{Udaipur Solar Observatory, Physical Research Laboratory, Udaipur 313 004, India}

\author{K. -S. Cho}
\affil{Korea Astronomy and Space Science Institute, Daejeon 305-348, Korea}

\author{Astrid M. Veronig}
\affil{IGAM/Institute of Physics, University of Graz, Universit$\ddot{a}$tsplatz 5, A-8010 Graz, Austria}







\begin{abstract}
We present RHESSI and TRACE observations of multiple flare activity that occurred in the active region NOAA 10656 over the period of two hours on 2004 August 18. Out of four successive flares, there were three events of class C while the final event was a major X1.8 solar eruptive flare. The events during the pre-eruption phase, i.e., before the X1.8 flare, are characterized by localized episodes of energy release occurring in the vicinity of an active region filament which produced intense heating along with non-thermal emission. A few minutes before the eruption, the filament undergoes an activation phase during which it slowly rises with a speed of $\sim$12~km~s$^{-1}$. The filament eruption is accompanied with an X1.8 flare during which multiple HXR bursts are observed up to 100--300 keV energies. We observe a bright and elongated coronal structure simultaneously in E(UV) and 50--100 keV HXR images underneath the expanding filament during the period of HXR bursts which provides strong evidence for ongoing magnetic reconnection. This phase is accompanied with very high plasma temperatures of $\sim$31 MK and followed by the detachment of the prominence from the solar source region. From the location, timing, strength, and spectrum of HXR emission, we conclude that the prominence eruption is driven by the distinct events of magnetic reconnection occurring in a current sheet formed below the erupting filament. These multi-wavelength observations also suggest that the localized magnetic reconnections associated with different evolutionary stages of the filament in the pre-eruption phase play a crucial role in destabilizing the filament by a tether-cutting process leading to large-scale eruption and X-class flare. 

\end{abstract}


\section{Introduction}
Solar eruptive phenomena correspond to various kind of transient activities observed in the solar atmosphere in the form of flares, prominence eruptions, and coronal mass ejections (CMEs). It is widely accepted that the fundamental processes responsible for these phenomena are closely related and of magnetic origin \citep{priest2002,lin2003}. There is also a near-universal consensus that magnetic reconnection is an essential aspect of the eruption and energy release processes. Investigation of the origin and propagation of solar eruptions is essential to further our understanding of the Sun-Earth connection.

Prominences, or equivalently filaments, are relatively cool, dense objects of chromospheric material suspended in the hotter corona by magnetic fields. When these structures erupt, both filament material and magnetic fields move out together. Decades of observations reveal that filament eruptions are often associated and physically related to CMEs and flares. Therefore study of prominence activity not only forms a very interesting topic of research in itself but also offers an opportunity to understand the physics of flares and CMEs. Filaments are commonly classified into two categories: active region and quiescent \citep{tandberg-hanssen1974,tang1987,tandberg-hanssen1995}. Active region filaments are low lying and rapidly evolving structures, forming in the newly emerged magnetic fields of an active region. They do not protrude much over the solar limb. On the other hand, quiescent prominences are generally associated with the decaying phase of an active region and are long lived. They tend to be located high in the corona. Some times activation and eruption phases of filaments are associated with flaring activity. In such cases, location, timing and, strength of X-ray emission provide important clues about crucial physical processes, such as, site of magnetic reconnection, particle acceleration, heating, etc. \citep{ding2003,ji2003,moon2004,alexander2006,liu2009,liuW2009,vema2012}.

Examination of subtle activity occurring in the vicinity of the filament and morphological changes within the filament during the pre-eruption phase has been considered very important to investigate the physical conditions of the solar atmosphere that lead to rapid energy release and large-scale eruption  \citep[e.g.,][]{farnik2003,chifor2007,joshi2011}. However, due to the limited sensitivity of the detector, a detailed X-ray imaging analysis during this phase of  mild energy release is often not possible. We still do not have a clear idea about the association between short-lived magnetic reconnection events occurring in the pre-eruption phase and the main flare that involve large-scale magnetic reorganization.  In this regard, it is very important to explore the pre-flare phase in order to determine the location/height of the reconnection sites as well as non-thermal characteristics of energy release. The two representative solar eruption models--tether-cutting and breakout--exploit the role of the initial magnetic reconnection in two different ways in order to set up the conditions favorable for the core fields to erupt. The "tether-cutting model" is fundamentally based on a single, highly sheared magnetic bipole, with the earliest reconnection occurring deep in the sheared core region \citep{moore2001}. On the other hand, in the "breakout model" the fundamental topology of the erupting system is multi-polar. Here the eruption is initiated by reconnection at a neutral point located in the corona, well above the core region \citep{antiochos1999}. In this manner, the former is built on the concept of an "internal reconnection" while the latter is suggestive of an "external reconnection" \citep{sterling2001}. By examining the observational signatures of magnetic reconnection during the pre-eruption phase, we can have clarity on the role of reconnection in triggering the eruption in the context of these two models.

In this paper, we present a comprehensive multi-wavelength analysis of 
multiple flare activity that occurred in active region NOAA 10656 on 2004 August 18. In total, we present observations of four successive flares. Out of these, three events are of class C which occurred before the main eruption in the vicinity of a filament, while the final X-class flare is associated with the filament eruption. The objective of this study is to understand the role of magnetic reconnection during various stages of the eruption. We also focus on the important morphological changes in the filament prior to the eruption. From this study, we expect to refine our knowledge on the triggering mechanism of solar eruptions. This investigation utilizes an excellent set of high resolution measurements from two unprecedented missions: Reuven Ramaty High Energy Solar Spectroscopic Imager (RHESSI) and Transition Region and Coronal Explorer (TRACE). In Section \ref{sec_obs_analysis}, we present multi-wavelength data analysis including X-ray spectroscopy. We discuss and interpret our observations in Section \ref{sec_discussion}. In Section \ref{sec_conclusions}, we summarize the conclusions of the present study.

\section{Observations and data analysis}
\label{sec_obs_analysis}

\subsection{Pre-eruption events and eruptive X-class flare}
\label{sec_overview}

\begin{figure}
\epsscale{.85}
\plotone{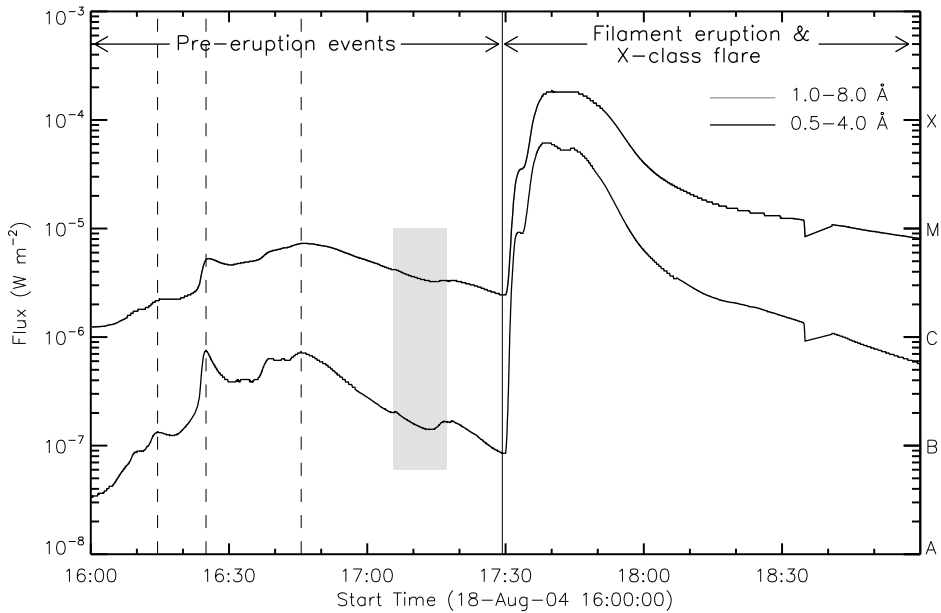}
\caption{Soft X-ray flux from the Sun observed by the GOES satellite in the 0.5--4 and 1--8~\AA~wavelength bands with a time cadence of 3 s between 16:00 and 19:00 UT. The vertical solid line differentiates the pre-eruption phase from the eruptive X1.8 flare.  The vertical dashed lines during the pre-flare interval indicate three subpeaks at 16:14:30, 16:25:00, and 16:45:39 UT during which EUV images reveal three localized instances of energy release (denoted as pre-events I, II, and III; cf. Table \ref{tab1}). 
The gray shaded region in the light curves correspond to the interval during which the filament exhibited fast morphological evolution.}
\label{goes_profile}
\end{figure}

In Figure \ref{goes_profile}, we show the soft X-ray flux from the Sun observed by  the GOES satellite in the 0.5--4 and 1--8~\AA~wavelength bands between 16:00 and 19:00 UT on 2004 August 18. From this light curve, one can clearly distinguish two phases of the flux evolution. The first phase extends from 16:00--17:30 UT and is characterized by a gradual rise and fall of the GOES flux. We note that the gradually varying SXR flux  during this phase is superimposed by three impulsive sub-peaks at 16:14:30, 16:25:00, and 16:45:39 UT and indicated in Figure \ref{goes_profile} by vertical dashed lines. A careful examination of the GOES flux with EUV images taken from TRACE (details in section \ref{sec_rhessi_trace}) reveals that these subpeaks represent localized events of energy release in the vicinity of a filament. 
Further we note that these three subpeaks are more pronounced in the high energy channel of GOES (i.e., 0.5--4~\AA~band) with the second one as the most impulsive. Hereafter, these subpeaks are termed as pre-events I, II, and III as they correspond to the energy release prior to the eruption. The second phase corresponds to a major eruptive flare of class X1.8 between 17:30 and 19:00 UT during which the filament erupts as a part of a coronal mass ejection. The multi-wavelength data set, analyzed in the following sections, suggest a link between the sequence of activities during the pre-eruption phase and subsequent large-scale eruption observed in the form of an X-class flare-CME event. 
The initiation and evolution of the associated CME are presented in \cite{cho09}. 

\subsection{Structure of active region}
\label{sec_active_region}

\begin{figure}
\epsscale{.85}
\plotone{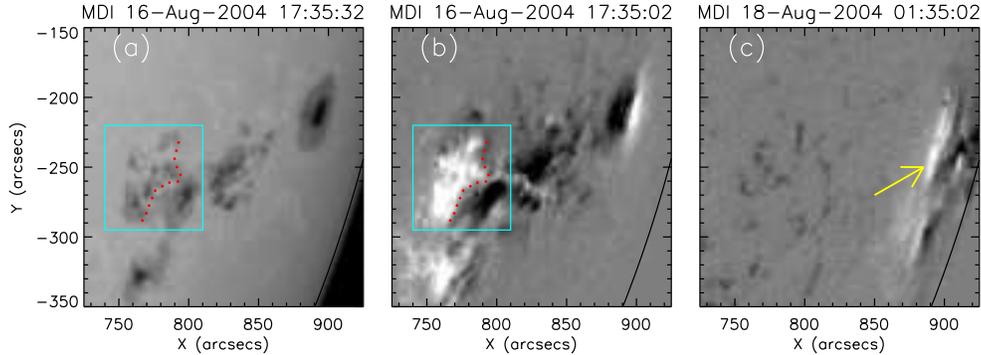}
\caption{SOHO/MDI observations of active region NOAA 10656. Since the active region was very close to the solar limb at the time of activity, we show a white light image and magnetogram taken two days prior to the event in panels (a) and (b) respectively. A close inspection of the region of interest (enclosed by a box) indicates a complicated magnetic polarity distribution on the surface. An approximate estimation of magnetic polarity inversion line is shown by dotted line. The active region on the day of the events under study is shown in panel (c) and the activity site is marked by an arrow.}
\label{mdi_int_mag}
\end{figure}

On 2004 August 18, solar active region NOAA 10656 was very close to the west limb of the Sun, with a mean position at S14 W90. The white light images taken from SOHO/MDI indicate that, at the time of the events, the leading part of the active region was behind the solar limb. Therefore, to get a better understanding of the magnetic configuration of the activity site, we present a white light image and a magnetogram taken by SOHO/MDI two days before the reported event in Figures~\ref{mdi_int_mag}(a) and (b). We note that the active region had a complex $\beta \gamma 
\delta$ magnetic configuration with sunspot clusters of negative and positive polarities as the leading and trailing sunspot groups respectively. The region of interest lies in the trailing part of the active region (shown inside a box in Figures~\ref{mdi_int_mag}(a) and (b)). We find that the active region is spatially complex. However, it is also apparent that the activity site does not exhibit a lot of mixing of polarities so that a simple polarity inversion line (PIL) can be defined (indicated in Figure~\ref{mdi_int_mag}(b) by red dotted line). The events occurred near the PIL and the activity site is marked by an arrow in Figure~\ref{mdi_int_mag}(c). The TRACE EUV images clearly show that a filament structure exists along the PIL (cf. Section~\ref{sec_rhessi_trace}).
From these longitudinal magnetograms, we conclude that the flaring region is mainly associated with a bipolar distribution of magnetic fields at the photosphere.  

\subsection{RHESSI X-ray light curves}
\label{sec_rhessi_lc}

\begin{figure}
\epsscale{.85}
\plotone{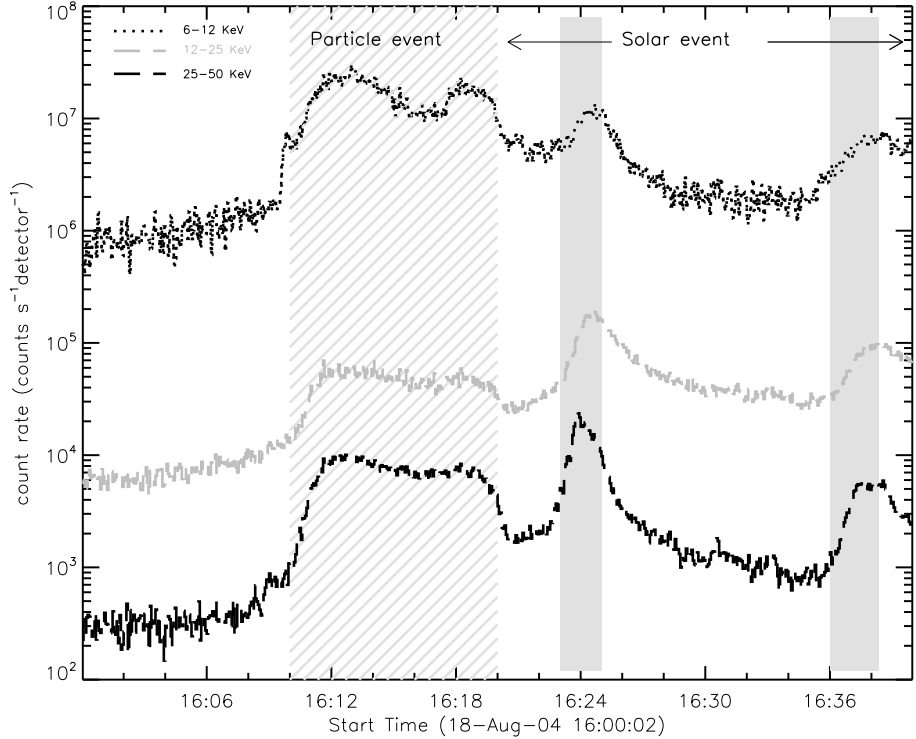}
\caption{RHESSI time profiles of the pre-eruption events with a time cadence of 4 s. The cross hatched region indicates the time intervals during which RHESSI observations are contaminated by a particle event. In order to present different RHESSI light curves with clarity, the RHESSI count rates are scaled by factors of 1/80, 1/5, and 1 for the energy bands 6-–12, 12-–25, and 25-–50 keV, respectively.  The gray shaded areas denote the time interval during which X-ray spectra were computed as shown in Figure~\ref{rhessi_spec_multiple_events}.}
\label{rhessi_lc_pre}
\end{figure}

\begin{figure}
\epsscale{.85}
\plotone{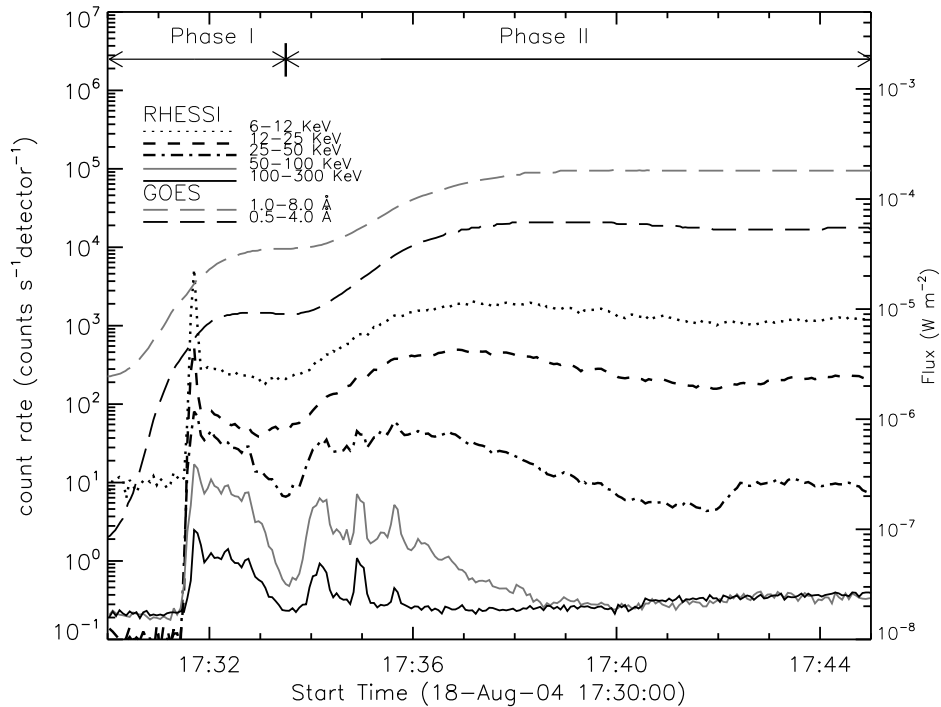}
\caption{RHESSI time profiles during the filament eruption and associated X1.8 flare with a time cadence of 4 s. In order to present different RHESSI light curves with clarity, the RHESSI count rates are scaled by factors of 1, 1/20, 1/30, 1/40 and 1/60 for the energy bands 6-–12, 12-–25, 25--50, 50--100, and 100-300 keV, respectively. For comparison, the soft X-ray flux in two wavelength channels observed by GOES are also plotted. We identify two phases in the evolution of X-ray flux. We note prolonged non-thermal emission during phase I while phase II is marked by three HXR bursts, which are observed up to 100--300 keV.}
\label{rhessi_lc_X1.8}
\end{figure}

RHESSI \citep{lin02} observations of the pre-eruption phase are available from 16:00 to 16:40 UT (Figure \ref{rhessi_lc_pre}). The light curves in different energy bands (viz., 6--12, 12--25, and 25--50 keV) are constructed by taking average count rates over front detectors 1, 3-–6, 8, and 9 in each energy band. We note that the RHESSI counts rates between 16:10 and 16:20 UT are contaminated by a particle event (i.e., the RHESSI detectors were hit by high-energy particles trapped in the Earth$^\prime$s radiation belts). This interval covers only the first subpeak of the GOES profile during which the level of soft X-ray flux was relatively low. The second subpeak, which is the most impulsive, was nicely covered by RHESSI up to 25--50 keV energy band. The third flare was partially observed by RHESSI between 16:35 and 16:40 UT as the spacecraft entered in the South Atlantic Anomaly (SAA) thereafter. 

In Figure~\ref{rhessi_lc_X1.8}, we present RHESSI and GOES X-ray light curves during the eruptive X1.8 flare. The examination of GOES profiles clearly indicate that the flare emission is associated with two distinct phases (marked in Figure~\ref{rhessi_lc_X1.8} as phase I and phase II). The first phase is characterized by a rapid rise and gradual decline of SXR flux between 17:30:00 and 17:33:30 UT. The flux further enhances after 17:33:30 UT and maximizes at $\sim$17:40 UT which marks the overall maximum of the event. It is noteworthy that the first phase is associated with intense and prolonged high energy emission up to 100-300 keV. On the other hand, HXR emission during the second phase occurred in the form of three distinct HXR bursts which are clearly identified in all the HXR channels above 25 keV.

\subsection{(E)UV and X-ray imaging}
\label{sec_rhessi_trace}

\begin{figure}
\epsscale{.85}
\vspace{-1.8cm}
\plotone{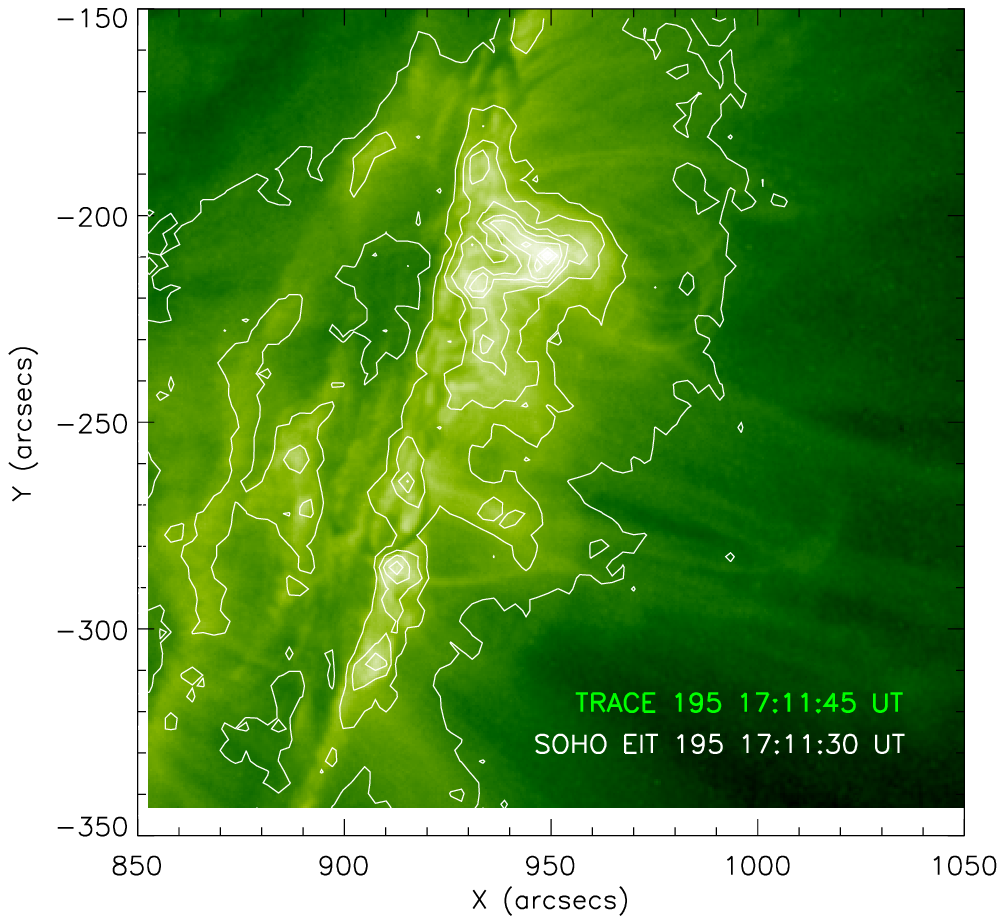}
\vspace{-0.8cm}
\caption{An illustration of the alignment of nearly co-temporal TRACE and SOHO/EIT 195~\AA~images. \vspace{0.25cm}}
\label{trace_align}
\end{figure}

The TRACE telescope has a field of view (FOV) of 8\arcmin.5 $\times$ 8\arcmin.5 and a spatial resolution of 1\arcsec (0\arcsec.5 per pixel). During the period of reported events, TRACE was monitoring the active region mostly with its EUV channel at 195~{\AA}. However, during the filament eruption TRACE provided a few UV images also  at 1600~{\AA} wavelength. The TRACE 195~{\AA} filter is mainly sensitive to plasmas at a temperature around 1.5 MK (Fe XII) but during flares it may also contain significant contributions of plasmas at temperatures around 15--20 MK \citep[due to an Fe XXIV line;][]{handy99}. TRACE 1600~{\AA}~channel is sensitive to plasma in the temperature range between (4--10)$\times$10$^{3}$ K and represents a combination of UV continuum, C I, and Fe II lines \citep{handy99}. The brightest and most rapidly varying features in the TRACE 1600~{\AA}~channel are likely to emit also in the C IV lines \citep{handy1998}.

To reconstruct RHESSI images, we have primarily used CLEAN algorithm \citep{hurford2002}. During the X1.8 flare, we also present X-ray images reconstructed by computationally expensive PIXON algorithm \citep{metcalf1996} to show precisely the location of X-ray emission during some of the crucial stages of the eruption. PIXON algorithm provides more accurate image photometry than CLEAN algorithm and is considered the best method to image extended sources in the presence of compact sources \citep{aschwanden2004}. The images are reconstructed by selecting front detector segments 3--8 (excluding 7) with 20 s integration time. 

Our analysis includes the identification of spatial distribution of the hard X-ray emission derived from RHESSI measurements with respect to the filament evolution observed in E(UV) wavelengths from TRACE. However, pointing information of TRACE is often not good enough to obtain the precise co-alignment between TRACE and RHESSI images. For a full disk imager (such as SOHO/EIT), the solar limb provides a reference which helps in accurate determination of the absolute pointing of the telescope. Moreover, thermal bending of the TRACE guide telescope also leads to an unknown variation in the pointing of at least a few arcseconds \citep{fletcher2001,alexander2006}. In order to correct the TRACE pointing, we have co-aligned near-simultaneous TRACE and SOHO/EIT images observed at same wavelength (i.e., 195 \AA) using the method of \cite{gallagher2002}\footnote{http://www.tcd.ie/Physics/people/Peter.Gallagher/trace-align/index.html}. 
This method is quite suitable for near-limb events where solar features in white light images do not appear very prominent \citep[see e.g.,][]{liu2009feb,bak2011}. We found that the TRACE pointing was offset by 2\arcsec.7 $\pm$0\arcsec.2 in the
X direction and 6\arcsec.8 $\pm$0\arcsec.3 in the Y direction (cf. Figure~\ref{trace_align}.)

\subsubsection{Pre-eruption events}

\begin{figure}
\epsscale{.82}
\plotone{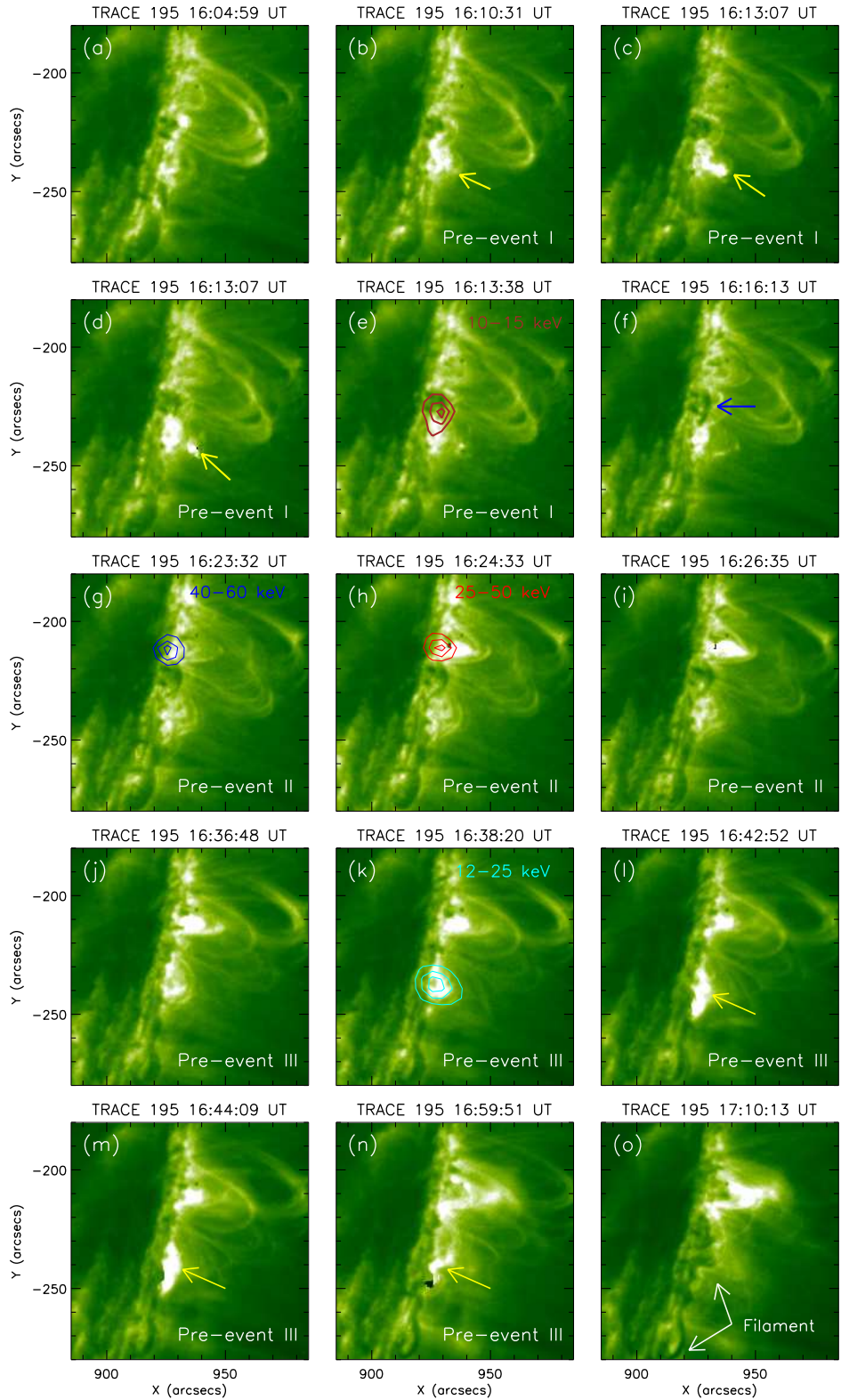}
\caption{Sequence of TRACE~195~{\AA} images showing three successive flares during the pre-eruption phase (i.e., pre-events I, II, and III) corresponding to the three subpeaks indicated in the GOES profile (cf. Figure \ref{goes_profile}). Note the intense brightening at the looptop (panel (b)) and formation and ejection of plasmoid (marked by arrows in panels (b)-(d)) which is accompanied with the pre-event I (panel (e)). Following pre-event I, the filament rises (indicated in panel (f) by black arrow). Pre-event II is very impulsive during which HXR sources up to 40-60 keV could be reconstructed (panels (g)-(i)). Pre-event III is marked by successive brightening in two low-lying loops (marked by arrows in panels (l)-(n)). RHESSI images are reconstructed with the CLEAN algorithm. The contour levels for RHESSI images are 70\%, 85\%, and 95\% of the peak flux in each image.} 
\label{trace_hessi_pre-event}
\end{figure}

The availability of TRACE 195~\AA~images at high time cadence ($\sim$30 s) during most of the pre-eruption phase (16:00 and 17:22 UT) enable us to examine the minute changes in the activity site. In Figure \ref{trace_hessi_pre-event}, we present a few representative TRACE 195~\AA~images. In the very beginning ($\sim$16:00 UT) the active region was relatively quite (cf. Figure \ref{trace_hessi_pre-event}(a)). We find that a bright system of loops existed in the northern part of the active region while its southern part lacked such coronal structures. Also it is noteworthy that the active region did not exhibit very large, complex overlying field lines in EUV images despite its extended structure. This indicates a relatively simplified coronal structure of the active region. 

After $\sim$16:07 UT, an intense brightening occurred at the southern side of the loop system (cf. Figure \ref{trace_hessi_pre-event}(b)) marking the onset of the first event of the pre-eruption phase (or `pre-event I'). The brightening was observed till 16:22 UT with the maximum SXR intensity up to C2.2 level at 16:14:30 UT. The flaring area grew in size rapidly and intense emission is produced at the looptop. At $\sim$16:12 UT, a blob like structure (i.e., plasmoid) was formed at the looptop which got detached from the flaring loops  (plasmoid is marked by arrows in Figures~\ref{trace_hessi_pre-event}(b)-(d)). The plasmoid showed an upward motion which can be clearly seen up to $\sim$16:15 UT. A very important feature that we note in this interval was the emergence of a filamentary structure (part of a long filament) just after the pre-event I (marked by an arrow in Figure~\ref{trace_hessi_pre-event}(f)). 
We also note that the event occurred at the southern leg of this rising filament. Although the RHESSI observations during this interval were contaminated by a particle event, we could still reconstruct 10--15 keV X-ray images at the peak time (i.e., 16:13:40 UT). The X-ray source is observed to be cospatial with intense flare loops (cf. Figure~\ref{trace_hessi_pre-event}(e)).

At $\sim$16:23 UT, we notice the start of pre-event II in the form of a sudden brightening near the second (i.e., northern) leg of the filament. The intensity and area of the flare showed a very fast rise and decay between 16:23--16:30 UT, implying an event of confined category. The GOES profile is very impulsive that peaked at 16:25:00 UT marking an event of C6.7 class, while the maximum flux of HXR emission was observed earlier (16:23:50 UT in 25--50 keV energy band).  From EUV images, we notice that there is a rapid expansion of height of the loop system in the corona during $\sim$16:23--16:27 UT. The volume as well as the intensity of the loop system displayed a fast decline thereafter. The RHESSI observations reveal HXR sources in 40--60 keV and 25--50 keV energy bands which are located near the base of an expanding loop system (cf. Figures~\ref{trace_hessi_pre-event}(g) and (h)).

The soft X-ray flux again strengthened after $\sim$16:35 UT and maximized at 16:45:39 UT which correspond to pre-event III of C7.3 class or overall maximum of the SXR profile during the pre-eruption phase (cf. Figures~\ref{trace_hessi_pre-event}(j)-(m)). This event occurred at the southern side of the filament and part of it is cospatial with the pre-event I.  Compared to two events described earlier, this event presented a very different morphological evolution. Initially (i.e., at $\sim$16:35) a single low-lying loop system brightened up and expanded. This is followed by the brightening of another adjacent small loop system at the southern side of existing flaring loops (marked by arrows in Figures~\ref{trace_hessi_pre-event}(l)-(n)). 
The coupled structure of two loop systems was visible up to $\sim$16:57 UT. These EUV observations are consistent with GOES profile that displayed broad maximum phase followed by gradual decline. RHESSI partially observed this phase up to $\sim$16:40 UT. We find that X-ray sources are observed in $\lesssim$25 keV energy bands from $\sim$16:36 UT onwards. The X-ray emission sources are spatially correlated with the flaring loops observed in EUV images (cf. Figure~\ref{trace_hessi_pre-event}(k)).

\begin{figure}
\epsscale{.85}
\plotone{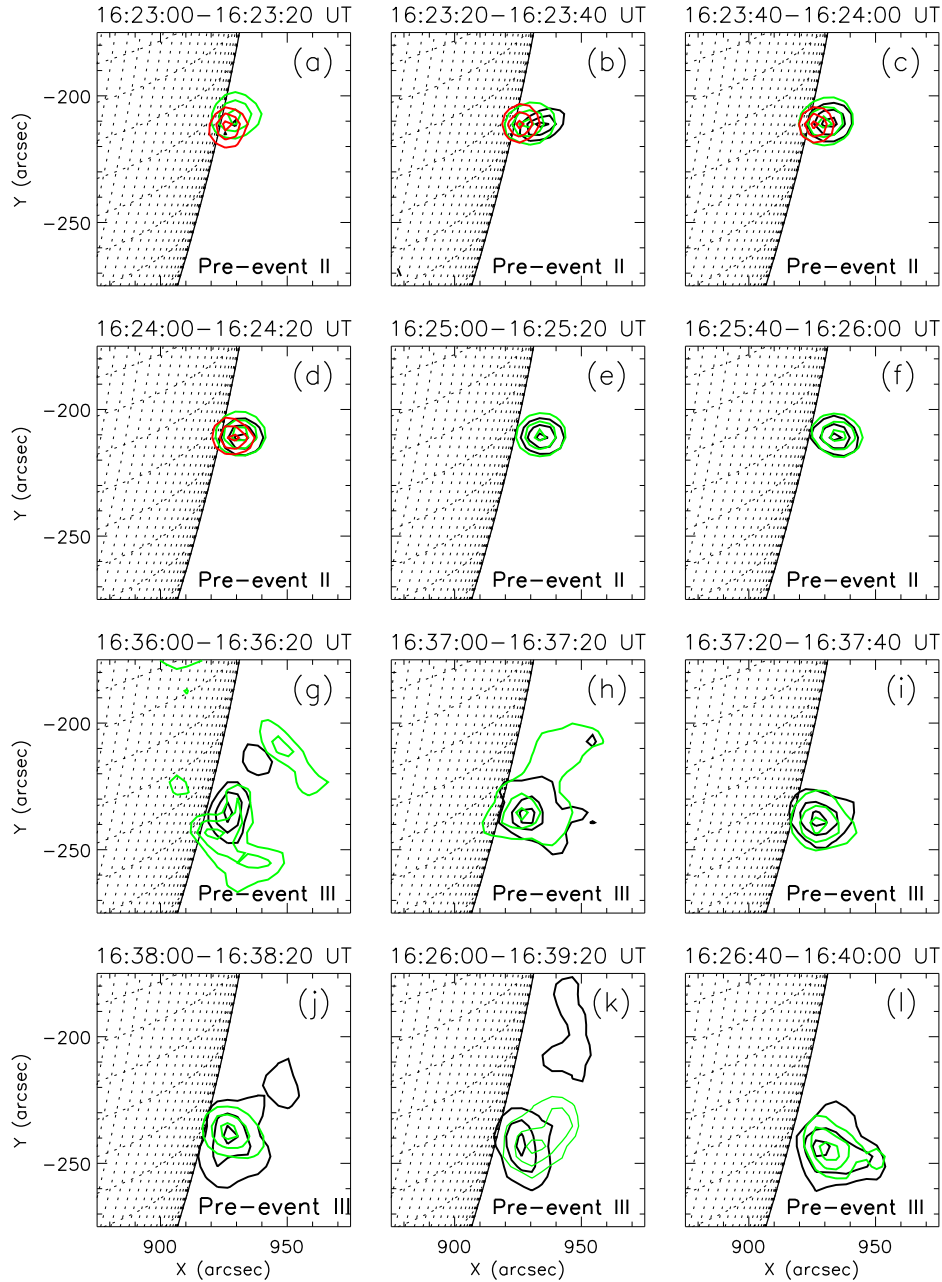}
\caption{Evolution of RHESSI X-ray sources during pre-events II and III in 6--12 keV, (black), 12--25 keV (green), and 25--50 keV (red) energy bands. RHESSI images are reconstructed with the CLEAN algorithm. The contour levels are 60\%, 80\%, and 95\% of the peak flux in each image.}
\label{rhessi_img_pre-event}
\end{figure}

In Figure~\ref{rhessi_img_pre-event}, we show the evolution of X-ray sources in different energy bands during the pre-events II and III. As discussed in sections~\ref{sec_overview} and \ref{sec_rhessi_lc}, pre-event II was very impulsive. It is noteworthy that the X-ray sources are compact in all the energy bands (i.e., 6--12 keV, 12--25 keV, and 25--50 keV) throughout this event (cf. first and second rows of Figure~\ref{rhessi_img_pre-event}). 
Due to the compactness of flaring region as well as proximity to the limb (see also Figures~\ref{trace_hessi_pre-event}(g)-(i)), it is hard to resolve the emission sources corresponding to the looptop and footpoint regions. 
However, in the decay phase (cf. Figures~\ref{rhessi_img_pre-event}(e) and (f)), when high energy HXR source vanishes, we can recognize distinct looptop emission in the form of upward moving 6--12 keV and 12--25 keV sources. In Figures~\ref{rhessi_img_pre-event}(g)-(l), we present RHESSI X-ray images in 6--12 keV and 12--25 keV energy bands in the early phase of pre-event III. In the beginning, the source structure in both energy bands is broad. Especially from 12--25 keV energy band images (cf. Figure~\ref{rhessi_img_pre-event}(g)), we can clearly see an extended source with two centroids that likely represent emission from the looptop and footpoint locations. The extended source structure also indicates presence of hot plasma in larger volume and is consistent with the structure of the flaring region observed in EUV images. In the later stages, both X-ray sources move upward with the high energy source (12--25 keV) always located at higher altitudes in comparison to low energy source (6--12 keV). 

\subsubsection{Activation of filament}
\begin{figure}
\epsscale{.9}
\plotone{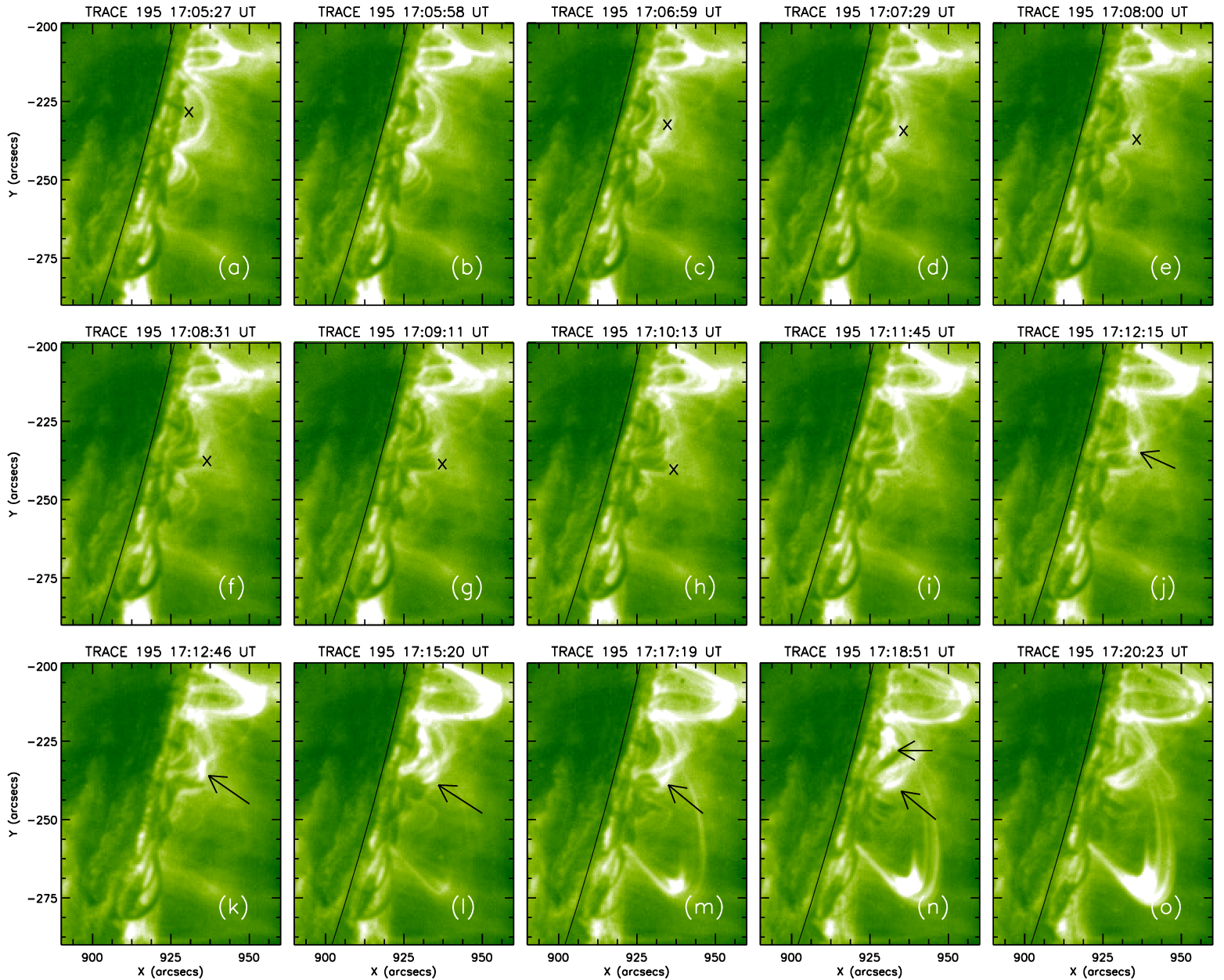}
\caption{Sequence of TRACE~195~{\AA} images showing the activation of filament after the pre-eruption events. The filament shows rapid evolution between 17:05 UT and 17:19 UT. This interval is marked by gray shaded region in Figure~\ref{goes_profile}. The rise of filament is shown by placing a cross (`$\times$') at the top. Note brightenings occurring below the apex of the prominence from $\sim$17:12 UT onwards (marked by arrows in panels (j)-(n)).}
\label{trace_195_filament}
\end{figure}

After the pre-eruption events, the filament undergoes a very important stage of morphological evolution. This phase was observed between 17:05 and 17:19 UT, i.e., just after the event III (indicated by gray shaded region in Figure \ref{goes_profile}). We present a sequence of TRACE 195~{\AA} images in order to show a closer and clearer view of filament activation in Figure~\ref{trace_195_filament}. The growth of filament is shown by placing a cross (`$\times$') at the top of the rising filament. From a linear fit to height-time data, we estimate speed of rising filament as $\sim$12~km~s$^{-1}$. 
After $\sim$17:11 UT, we observe a localized brightening (indicated by arrows in Figures \ref{trace_195_filament}(j)-(n)) which is located at the top of a twisted flux rope, but below the apex of the filament. This brightening grows in successive images. We note another localized brightening below the rising filament at $\sim$17:18 UT (indicated by another arrow in Figure \ref{trace_195_filament}(n)). Due to unavailability of TRACE observations between 17:22 and 17:34 UT, we are unable to track the later phase of filament activation. RHESSI observations during this phase are not available.

\subsubsection{Filament eruption and X1.8 flare}
\label{sec_trace_hessi_x1.8}
\begin{figure}
\epsscale{.85}
\plotone{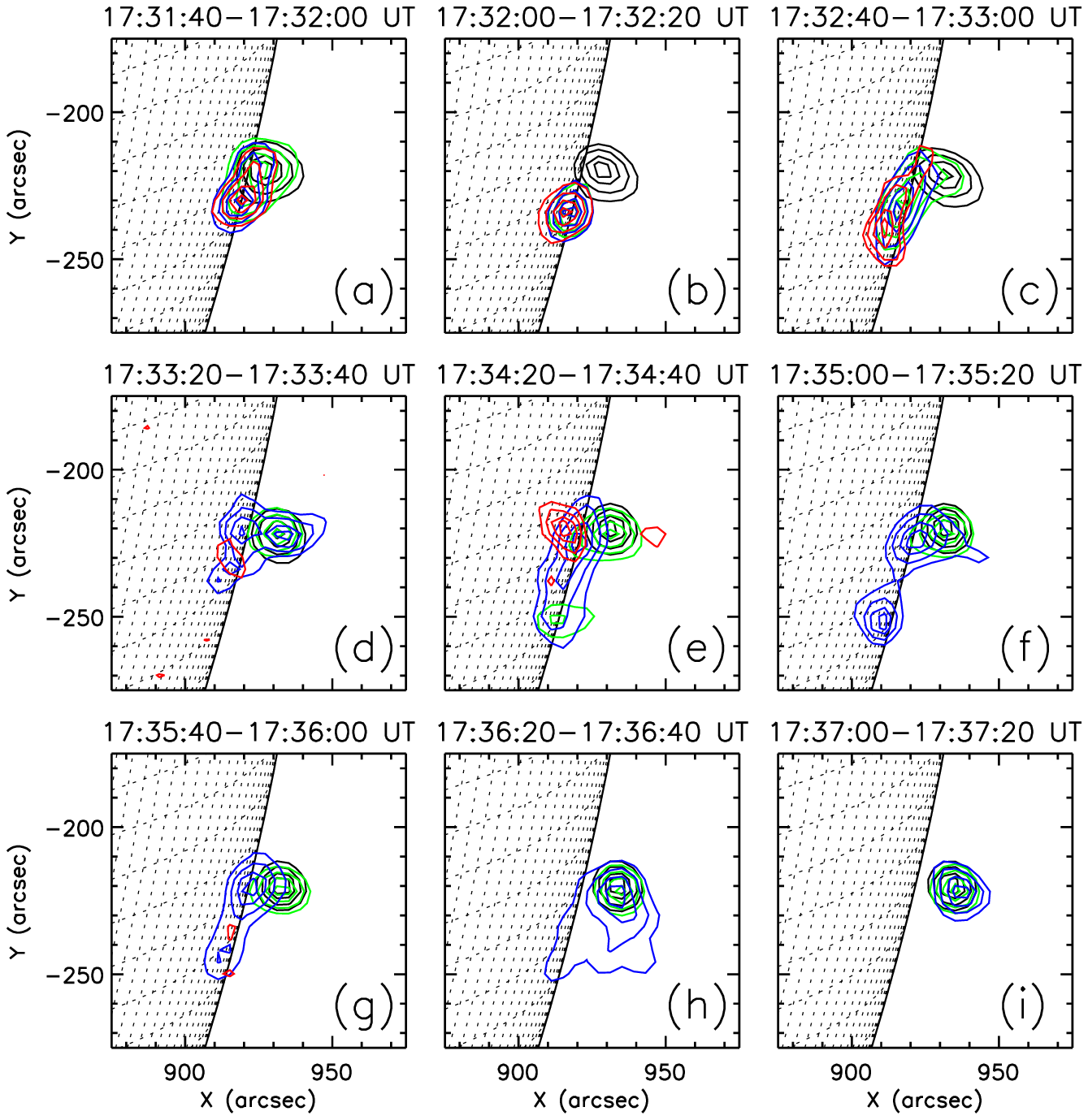}
\caption{Evolution of RHESSI X-ray sources during eruptive X1.8 flare (cf. Figure~\ref{goes_profile}) in 12--25 keV (black), 25--50 keV (green), 50--100 keV (blue), and 100--300 keV (red) energy bands. RHESSI images are reconstructed with the CLEAN algorithm. The contour levels are 55\%, 70\%, 85\%, and 95\% of the peak flux in each image.}
\label{rhessi_img_X1.8}
\end{figure}

\begin{figure}
\epsscale{.90}
\plotone{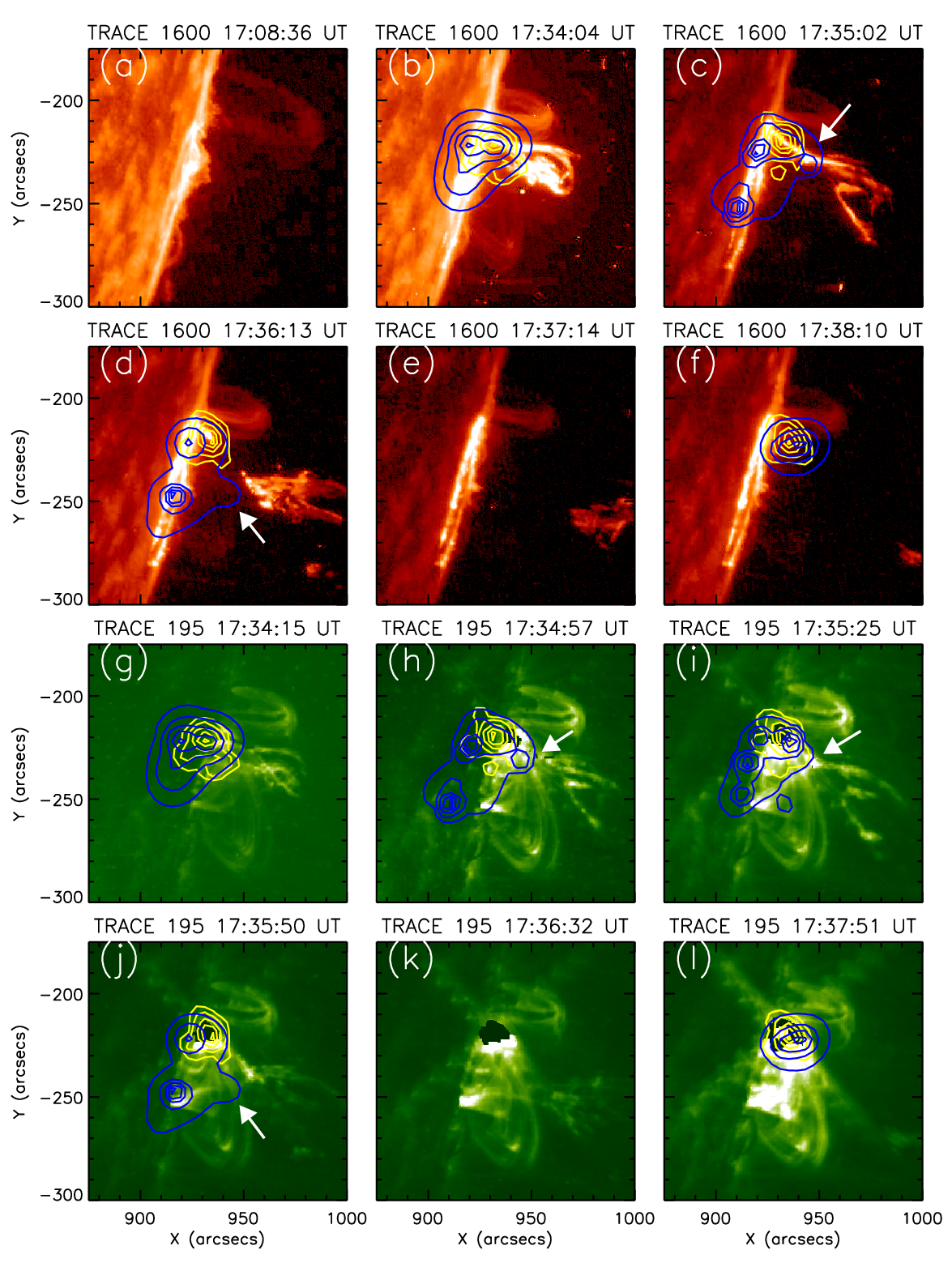}
\caption{A few representative images taken by TRACE in 1600~{\AA} and  195~{\AA}  channels during eruptive X1.8 flare (cf. Figure~\ref{goes_profile}) overlaid by co-temporal RHESSI X-ray images in 12--25 keV (yellow) and 50--100 keV (blue) energy bands. It is noteworthy that 50-100 keV HXR coronal source lies over an elongated, bright (E)UV structure formed below the erupting prominence (marked by arrows in panels (c), (h) and (i)). We also note 50-100 keV HXR emission from an extended region in the corona during the phase of detachment of the prominence from the solar source region (marked by arrows in panels (d) and (j)). The RHESSI images are reconstructed with PIXON algorithm. The contour levels for RHESSI images are 10\%, 25\%, 50\%, 75\%, and 95\% of the peak flux in each image.
}
\label{trace_hessi_X1.8}
\end{figure}

TRACE observations of the event are available after 17:34 UT, i.e., $\sim$4 minutes after the onset of the flare (GOES start time $\sim$17:30 UT). By this time the eruption of filament had already begun (cf. Figures~\ref{trace_hessi_X1.8}(b) and (h)). However, RHESSI observations are available from the beginning of the event (from ~17:31 UT onward). In Figure~\ref{rhessi_img_X1.8}, we present evolution of HXR sources in four energy bands (i.e., 12--25 keV, 25--50 keV, 50--100 keV, and 100-300 keV). We note that the HXR emission at 12-25 keV (black contours) is associated with a compact, single coronal source throughout the flare which also exhibits 
``standard" upward movement. The HXR sources at higher energies ($>$ 25 keV) reveal complicated evolution. We observe onset of very high energy HXR emissions (25--50 keV, 50--100 keV, and 100--300 keV) from the very early stages (Figures~\ref{rhessi_img_X1.8}(a)-(c)) which are mainly from the footpoint regions. This is followed by a very important phase during which high-energy coronal HXR sources are observed. In particular, there are instances when HXR emission up to 50-100 keV becomes very prominent from coronal regions besides footpoint locations (cf. Figures~\ref{rhessi_img_X1.8}(d) and (f)). We further note that at these times, coronal HXR emission is from an extended region. During the later stages, only coronal HXR emission is detected (Figures~\ref{rhessi_img_X1.8}(h) and (i)). We also note that the 50-100 keV HXR coronal source was observed until 17:38 UT. In the post flare phase, HXR coronal sources are observed only up to 25 keV energies. 
  
In Figure \ref{trace_hessi_X1.8}, we present a few representative TRACE 1600~\AA~(panels (a)-(f)) and 195~\AA~(panels (g)-(l)) images. We have overplotted co-temporal RHESSI X-ray PIXON images at 12--25 keV and 50-100 keV energy bands in selected panels to compare the location of HXR coronal and footpoint sources with respect to the phases of erupting filament. With the start of TRACE observations at 17:34, we observe expansion and stretching of a flux rope which rapidly evolved into a  `Y-shaped' structure  (cf. Figures~\ref{trace_hessi_X1.8}(b)-(c)). We observe intense brightening along the leg of the erupting structure in the form of an elongated bright structure (cf. Figure~\ref{trace_hessi_X1.8}(c)). This bright feature is also seen in 195~\AA~images (cf. Figures~\ref{trace_hessi_X1.8}(h) and (i)). It is noteworthy that the HXR coronal sources at 50--100 keV energy band lie over this   elongated bright (E)UV structure (marked by arrow in Figures~\ref{trace_hessi_X1.8}(c) and (h)). In the next frames during 17:35--17:36 UT (cf. Figures~\ref{trace_hessi_X1.8}(d) and (j)), we find the detachment of the stretched flux rope from the solar source region and its upward propagation. From a linear fit to height-time measurements of the erupting filament observed at 1600~{\AA} images, we estimate the speed of erupting filament as $\sim$270~km~s$^{-1}$. We note that during the detachment of the fluxrope, HXR emission at 50--100 keV is observed from an extended source covering footpoint as well as coronal regions (cf. Figures~\ref{trace_hessi_X1.8}(d) and (i)). During this phase, RHESSI light curves indicate three consecutive HXR bursts (cf. Figure~\ref{rhessi_lc_X1.8}). After the eruption, we observe a closed post-flare loop system with HXR emission from the coronal loops (cf. Figures~\ref{trace_hessi_X1.8}(e)-(f) and (k)-(l)).

\subsection{RHESSI X-ray spectroscopy}

\begin{figure}
\epsscale{.8}
\plotone{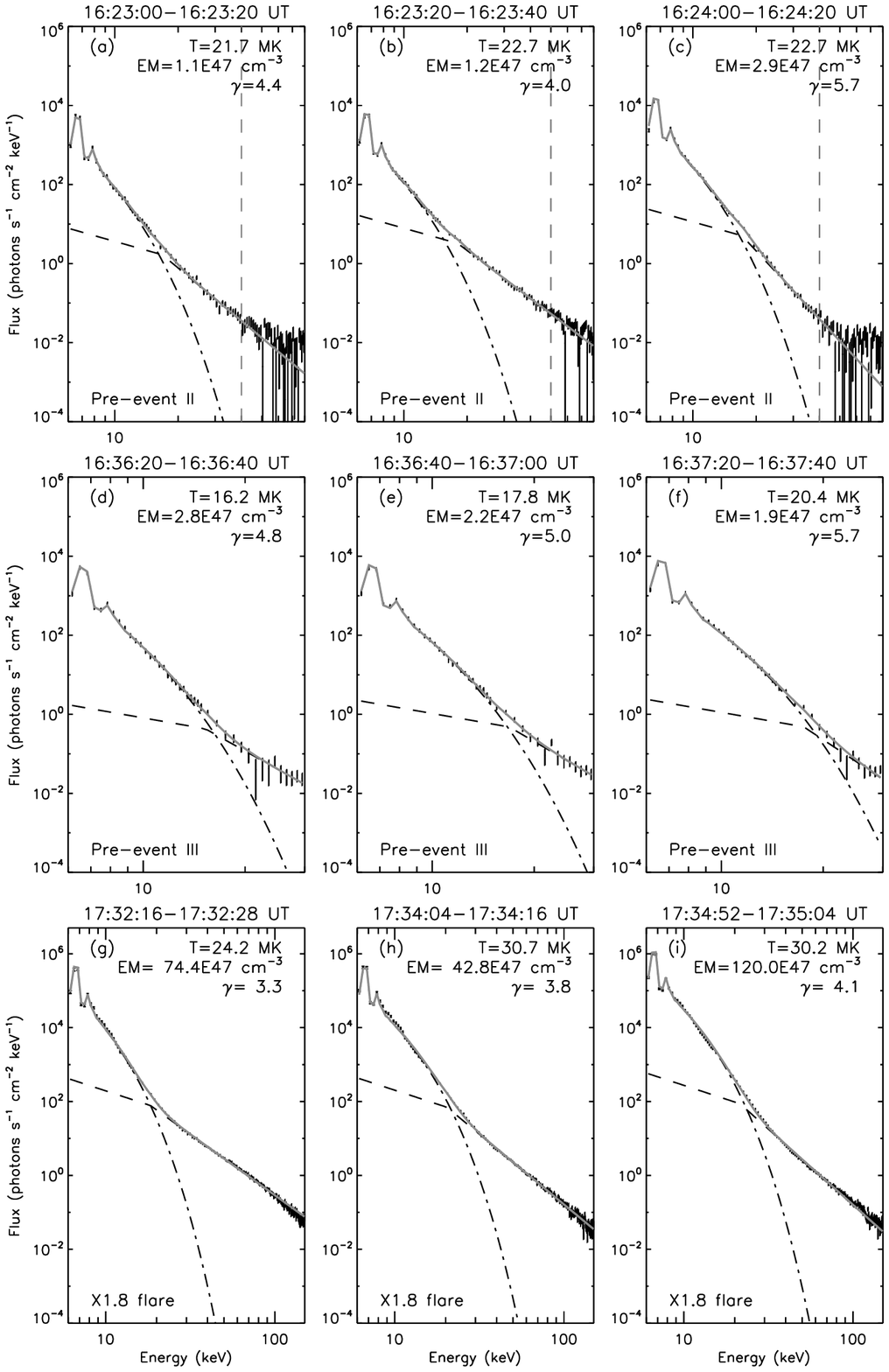}
\caption{RHESSI X-ray spectra derived during various time intervals covering the pre-eruption events II and III (first and second rows respectively) and X1.8 eruptive flare (last row) together with applied fits. The spectra were fitted with an isothermal model (dashed-dotted line) and a functional power law with a turnover at low energies (dashed line). The gray (solid) line indicates the sum of the two components. The maximum energy used for fitting each spectrum during pre-event II is denoted by a vertical dashed line (in top row only).}
\label{rhessi_spec_multiple_events}
\end{figure}

\begin{figure}
\epsscale{.85}
\plotone{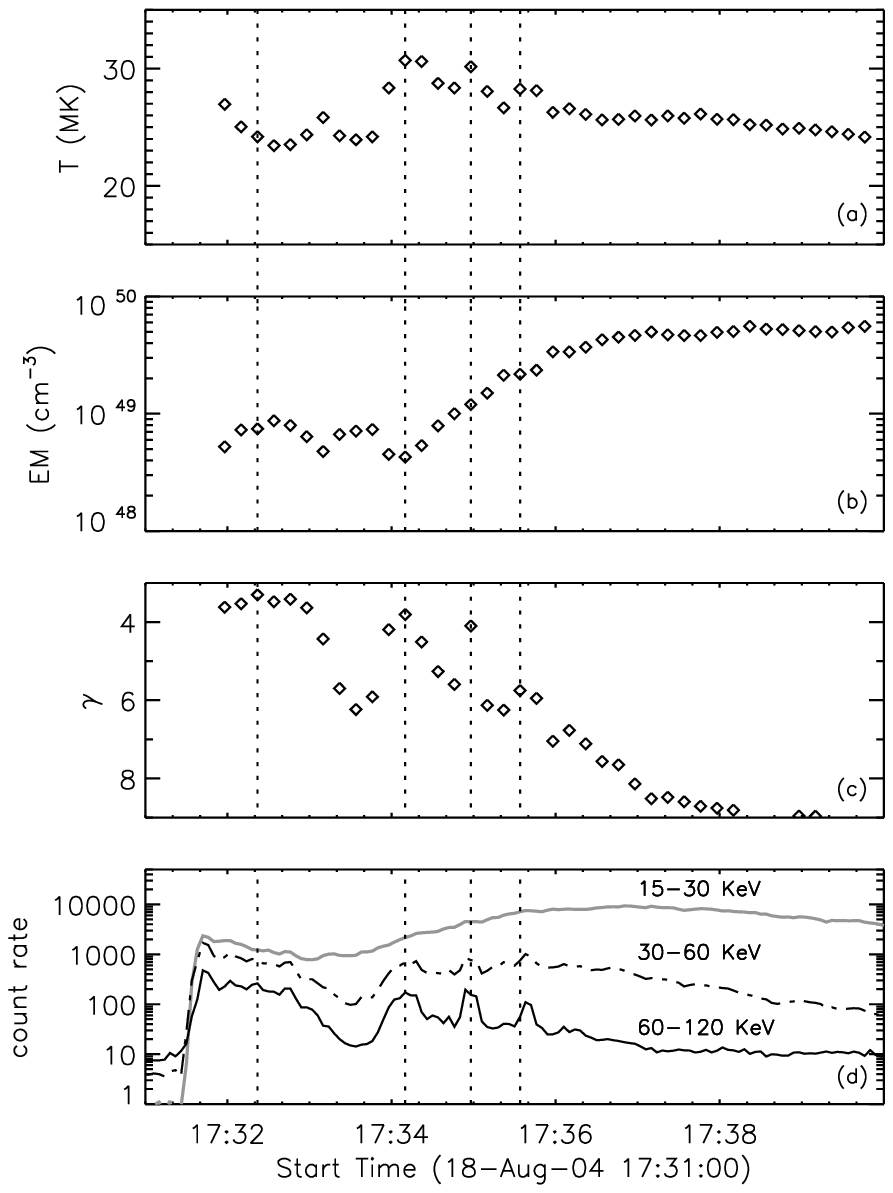}
\caption{Temporal evolution of various spectroscopic quantities derived
from RHESSI X-ray spectral fits of consecutive 12 s integration times for X1.8 flare. From top to bottom: plasma temperature, emission measure, photon spectral index, and RHESSI count rates in the 15--30 keV, 30--60 keV, and 60--120 keV energy bands. Dashed vertical lines indicate important phases of flare evolution during which spectra showed strong non-thermal characteristics.}
\label{rhessi_spec_param_X1.8}
\end{figure}

We have studied the evolution of RHESSI X-ray spectra
during the pre-eruption phase as well as the eruptive X1.8 flare. We first generated a RHESSI spectrogram with an energy binning of 1/3 keV for energy range of 6–-15 keV
and 1 keV for energies $>$15 keV. We only used the front segments of
the detectors, and excluded detectors 2 and 7 (which have lower
energy resolution and high threshold energies, respectively).
The spectra were deconvolved with the full detector response
matrix (i.e., off-diagonal elements were included;~\cite{smith2002}). Spectral fits were obtained using a forward-fitting method
implemented in the OSPEX code. OSPEX allows the user
to choose a model photon spectrum, which is multiplied with
the instrument response matrix and then fitted to the observed
count spectrum. The best-fit parameters are obtained as output.
We used the bremsstrahlung and line spectrum of an isothermal
plasma and a power-law function with a turnover at low
X-ray energies. The negative power-law index below the low-energy
turnover was fixed at 1.5. In this manner, there are five
free parameters in the model: temperature (T) and emission
measure (EM) for the thermal component, and power-law index
($\gamma$), normalization of the power law, and low-energy turnover
for the non-thermal component. From these fits, we derive the
temperature and EM of the hot flaring plasma as well as the
power law index for the non-thermal component. 

We computed spatially integrated, background subtracted RHESSI spectra accumulated for 20 s integration time during pre-events II and III (this time interval is indicated by gray colored strips in Figure~\ref{rhessi_lc_pre}). A few representative spectra during this interval are shown in Figures~\ref{rhessi_spec_multiple_events}(a)-(f). We find that during pre-event II, the X-ray emission exhibited a very fast spectral evolution with the peak HXR emission between 16:23 and 16:24 UT.
Due to fast spectral evolution of the flare we decided to select the maximum energy value for fitting by comparing the flux during the flare with that of the background level for each time interval. This can be easily achieved by selecting the fitting option Auto-Set-Max in OSPEX software. The maximum value of energy used for fitting is indicated by a dashed line in each spectrum shown in Figures~\ref{rhessi_spec_multiple_events}(a)-(c).  The lower energy value for fitting was selected as 6 keV. In Figures~\ref{rhessi_spec_multiple_events}(d)-(f), we show  a few representative RHESSI spectra derived during the early part of the event III. 
The spectra were fitted in the energy range of 6--30 keV. In Figures~\ref{rhessi_spec_multiple_events}(g)-(i), we show a few representative RHESSI spectra derived during X1.8 flare. The spectra were fitted in the energy range of 6--150 keV. Time evolution of various spectral parameters (temperature (T), emission measure (EM), and power-law index ($\gamma$)) obtained from fits to the RHESSI spectra, integrated over consecutive 12~s intervals, is shown is Figure~\ref{rhessi_spec_param_X1.8}. 

\section{Results and discussion}
\label{sec_discussion}

We present a multi-wavelength study of four successive flares that occurred in active region NOAA 10656 on 18 August 2004 over a period of two hours. Three flares of class C occurred before the eruption, while the last event was a major X1.8 eruptive flare. The pre-eruption events occurred in the vicinity of a filament and were localized in nature. The filament became unstable and eventually erupted, producing an X-class flare. Table \ref{tab1} presents an observational summary of flare activity and associated phenomena. We discuss the multiple flare activity in terms of two evolutionary stages:

\begin{table}
\begin{center}
\caption{Observational summary of activities during pre-eruption phase and eruptive X1.8 flare}
\begin{tabular}{p{1.2in}p{0.8in}p{1.8in}p{1.6in}}
\tableline\tableline\\
Phases & GOES interval in UT (peak time)  & \multicolumn{2}{c} {Activity}\\
\cline{3-4}\\
& & \multicolumn{1}{c}{TRACE (EUV)}&\multicolumn{1}{c}{RHESSI}\\ 
\tableline \\

Pre-event~\small{I} (C2.2)& 16:07--16:22 UT (16:14 UT)& Localized brightening along with plasmoid ejection at southern leg of the filament, filament slowly rises & Contaminated by particle event, compact source in 10--15 keV\\
\\
Pre-event~\small{II} (C6.7)& 16:23--16:30 UT (16:25 UT) & Confined flare at the northern leg of the filament &HXR source up to 50--80 keV detected with significant non-thermal emission ($\gamma = \sim$4.0)\\
\\
Pre-event~\small{III} (C7.3) & 16:35--17:05 UT (16:45 UT) & Sequential brightening in two low-lying loops followed by arcade formation & Partially observed up to $\sim$16:40 UT, mostly thermal emission from the looptop up to $\sim$30 keV \\
\\
Filament activation & 17:05--17:19 UT& Rise of filament along with localized brightenings &Not available\\
\\
Filament eruption and X1.8 flare & 17:30--19:00 UT&Filament eruption, formation of bright elongated coronal structure below the erupting prominence& Multiple HXR bursts with strong non-thermal HXR emission, high-energy coronal HXR sources \\
\tableline
\label{tab1}
\end{tabular}
\end{center}
\end{table}

\subsection{Pre-eruption phase}

The pre-eruption phase is characterized by three localized flares (i.e., pre-events I, II and III) along with heating, activation, and rise of the filament.
Pre-event I exhibited intense brightening of a low-lying loop system observed within in a very confined volume (see Figures~\ref{trace_hessi_pre-event}(b)-(e)). However, this small flare is remarkable in that it was associated with a plasmoid ejection. The formation of the plasmoid at the top of the flare loops and its upward motion is readily visible from EUV images. Another significant observation is the presence of 10--15 keV hard X-ray LT source at the flare maximum indicating intense plasma heating. The upward moving blob of hot plasma along with HXR LT source provide evidence for the magnetic reconnection. It is likely that ejection of plasma and magnetic fields, observed as plasmoid, are caused due to weaker overlying magnetic fields and lower density above the flare loops. Plasmoid ejection is believed to be a consequence of magnetic reconnection. We emphasize that the rise of the filament began after this small flare which occurred at its southern leg, suggesting a causal relation between the flare event and filament rise. It is likely that  the magnetic reconnection caused a weakening of overlying field lines making the expansion of the filament possible. 

Pre-event II corresponds to the very rapid variation of GOES flux that lasted only for $\sim$7 minutes. This highly impulsive and compact event took place at the northern leg of the filament (see Figures~\ref{trace_hessi_pre-event}(g)-(i)). Despite the impulsiveness of the event, we clearly identify that the peak of the HXR flux (16:23:50 UT in 25--50 keV energy band) occurred before the thermal soft X-ray peak (16:25:00 UT in GOES 1--8~\AA). The timings of peak of HXR and SXR flux suggest that high energy HXR emission is associated with the processes that are intimately linked with the primary energy release (i.e., magnetic reconnection) while the SXR flux with delayed maximum is attributed to the thermal emission from hot flare loops. The series of cospatial EUV and X-ray images readily confirm this picture. The EUV images reveal fast expansion of a loop system within a confined yet elongated region. We note that HXR emission was clearly observed up to 60 keV above the background level. It is noteworthy that X-ray spectra during the peak timings indicate hot thermal plasma ($\sim$23 MK) with lower values of emission measure (see Figures~\ref{rhessi_spec_multiple_events}(a)-(c)), indicating intense plasma heating within a confined environment \citep{joshi2011}. Further at this time the HXR spectrum reveals strong non-thermal characteristics at energies $>$15 keV with photon spectral index $\gamma$=$\sim$4. It is remarkable that in such a short-lived and confined event, we clearly recognize {\it{soft-hard-soft}} spectral evolution, providing evidence for particle acceleration \citep{benz1977,grigis2004,joshi2011}.

Compared to the previous events, Pre-event III showed a very different morphological evolution.  The TRACE EUV observations suggest that this event is associated with the successive brightening of two systems of low-lying loops which are located side by side (see Figures~\ref{trace_hessi_pre-event}(j)-(n)). 
The partially available RHESSI observations indicate the emission from coronal loops along with an upward movement of the LT source (Figure~\ref{rhessi_img_pre-event}), an important feature of standard flare model \citep[see e.g.,][]{joshi2007}. RHESSI X-ray spectroscopy analysis reveals lower values of temperature and emission measure during this event compared to pre-event II (Figure~\ref{rhessi_spec_multiple_events}).

During the time interval between event I and III, the evolution of filament was very slow. However, we emphasize that the filament's slow evolution in the pre-eruption phase is temporally and spatially associated with flaring activity. We observe more significant morphological changes associated with the rising filament after the pre-eruption events (Figure~\ref{trace_195_filament}). The changes started to occur just after event III and were continuously observed during the rest of the pre-eruption phase. During this interval the filament exhibited continuous rise with a speed of $\sim$12~km~s$^{-1}$. Before the main eruption, we find brightenings within the twisted filament channel which is also reflected in GOES observations as a small bump at $\sim$17:18 UT 
(Figure~\ref{goes_profile}). The localized brightenings at two different locations below the apex of the prominence is likely caused due to heating by magnetic reconnection \citep{chifor2007}. We further note that as the brightening increased, the filament appeared less structured probably due to the change from absorption of EUV radiation to emission caused by fast heating of plasma within the filament \citep{filippov2002}. 

\subsection{Filament eruption and X1.8 flare}
This phase is marked by the onset of X1.8 flare during which the activated filament underwent a transition into the dynamic phase and erupted. This phase is characterized by a fast rise of the prominence ($\sim$270~km~s$^{-1}$) and strong HXR non-thermal emission. 

The evolution of HXR flux during the X-class flare is very interesting. It displayed four distinct episodes of flux enhancement; the first peak was very broad (indicated as phase I; cf. Figure~\ref{rhessi_lc_X1.8}) and gradual while other three peaks were impulsive (indicated as phase II; cf. Figure~\ref{rhessi_lc_X1.8}). More importantly, these peaks were observed at very high energies, viz., up to $\sim$100--300 keV. During the phase I of prolonged HXR emission (17:31 UT-17:33 UT), the X-ray emission above $\sim$20 keV follows hard power laws with photon spectral index $\sim$3.5 (Figure~{\ref{rhessi_spec_param_X1.8}). The spectra continued to be harder during this whole phase. The spectra again became harder during the next three hard X-ray bursts (indicated by vertical dashed lines in Figure~\ref{rhessi_spec_param_X1.8}). 
We note a very important phase of prominence eruption in E(UV) observation between the second and third HXR peaks (at $\sim$17:35 UT). With the rapid expansion of the prominence, intense brightening is observed in (E)UV images below the erupting structure (Figures~\ref{trace_hessi_X1.8}(c) and (h)). It is remarkable that this important stage of prominence eruption is spatially and temporally associated with high energy HXR emission in the form of an extended coronal HXR source observed in 50--100 keV energy band. We further note that the temperature rose to very high values (with maximum value as $\sim$31 MK at $\sim$17:34 UT) during the HXR bursts occurred between 17:34$-$17:36 UT. 
This period of high plasma temperature matches with the appearance of an elongated bright (E)UV structure and extended HXR coronal source. This stage of multiple HXR bursts concluded with the final detachment of the prominence from the solar source region, with the closed post-flare loops remained on the solar limb. We also note that the thermal emission dominates in the decay phase with a slow rise in emission measure.

After the first HXR burst ($\sim$17:32-17:33), we observe strong HXR coronal emission at energies $\geq$25 keV (Figures~\ref{rhessi_img_X1.8}(d)-(i)) which was detected till late phases of the flare. Moreover, HXR images at 50-100 keV clearly indicate coronal emission between 17:35 and 17:38 UT. Such a high energy coronal HXR emission has recently been detected in several flares by RHESSI during their different evolutionary phases \citep{veronig2004,krucker2008,joshi2009,joshi2011}.
However, the physical mechanism for this strong non-thermal
source in the tenuous corona is still not clearly understood.

\section{Summary and conclusions}
\label{sec_conclusions}

Solar eruptions are complex phenomena which involve a diversity of physical processes during their various stages. The observations of multiple flare activity from AR 10656 on 2004 August 18 provide us with a unique opportunity to understand some of the aspects of the eruption process, viz., role of pre-eruption magnetic reconnection, triggering mechanism as well as particle acceleration and reconnection during the main phase of the eruption, which are summarized as follows:
 
{\bf 1.}  Our observations imply that pre-eruption events essentially represent discrete episodes of magnetic reconnection which played important role in filament evolution toward eruption. The evidence for 
localized magnetic reconnection and particle acceleration is observed in the form of plasmoid ejection, HXR emission and {\it soft-hard-soft} evolution of HXR spectra. The sequence of activities suggest that the localized and short-lived episodes of magnetic reconnection that occurred in the vicinity of the filament, would tend to weaken the overlying magnetic structure.  Here it is important to note that the decrease of overlying magnetic field is a crucial factor that permits the process of successful eruption of an unstable flux rope \citep{torok2004,torok2005}. Observations presented here reveal 
an early quasi-static, slowly evolving phase of the filament before the onset of more dynamic activation phase. The activation phase of filament associated with slow rise and heating has been recognized as an important precursor before the eruption \citep{sterling2005,liu2008,sterling2011}.

{\bf 2.} These observations also reveal that initiation of the eruption is essentially linked with the EUV/X-ray emissions originating at the lower corona and/or chromospheric heights rather than reconnection occurring in higher coronal loops, above the prominence. The HXR emission also suggests that the pre-eruption reconnection occurred close to the leg of the prominence. The heating of the rising prominence during its activation phase essentially resembles the precursor phase brightenings that are generally observed a few to tens of minutes prior to a large flare in the form of enhancements of soft X-ray and EUV emissions \citep{magara1999}. From these observations we infer that the mechanism leading to pre-eruption flares and precursor emission caused the onset of filament activation and eruption. Our present interpretation of the initiation of the prominence eruption is consistent with the tether-cutting mechanism \citep{moore1992}. The onset of the X1.8 flare marks the fast rise phase of the filament. \cite{sterling2005} suggested that the transition to the fast-rise phase would occur when the low-lying reconnection that initiated the slow-rise phase inflated the overlying filament-carrying fields enough that they became unstable and violently erupted, perhaps due to an MHD instability or due to runaway tether cutting.

{\bf 3.} The filament eruption was accompanied by an X1.8 flare. 
The flare is marked by four distinct HXR peaks up to 100--300 keV energies. 
We observed strong and prolonged non-thermal HXR emission right at the flare onset, evidencing an extended phase of particle acceleration \citep[see e.g., ][and references therein]{joshi2012} related to the early stages of 
filament eruption. We therefore interpret that the initial ejection of filament is associated with the formation of a current sheet underneath which then reconnects to cause the subsequent eruption and non-thermal emission \citep[see e.g.,][]{alexander2006}. RHESSI and TRACE observations during the next three HXR bursts are consistent with this interpretation which reveal a thin, elongated, and bright structure in E(UV) images cospatial with 50--100 keV extended coronal HXR sources. Within the scope of standard flare model, we believe that the extended coronal HXR emission associated with bright, thin (E)UV structure is a direct consequence of magnetic reconnection in the current sheet formed below the erupting prominence. This is further supported by the fact that at this very interval, the temperature attained the highest value of $\sim$31 MK. These three impulsive HXR bursts clearly showed {\it soft-hard-soft} spectral evolution suggesting distinct events of particle acceleration associated with the early phase of CME initiation. The appearance of strong HXR coronal as well as footpoint emissions during the impulsive phase imply rapid dissipation of magnetic energy in the current sheet as a result of increase in the rate of magnetic reconnection \citep[see e.g.,][]{sui2003}.

Understanding the triggering mechanism of solar eruptions is still a challenging topic of research in contemporary solar physics. The investigations of filament evolution in the pre-eruption phase and associated small-scale reconnection events could provide special insights into the processes that lead to the initiation of coronal mass ejections. We are in the process of analyzing suitable data with superior resolution observed at several E(UV) channels from the Solar Dynamics Observatory (SDO) to learn about the multiple stages of the eruptive phenomena more precisely. Probing the thermal and non-thermal characteristics of the relatively mild X-ray emissions during the pre-eruption phase is very crucial for such studies. We hope that the observation of the proposed Solar Low-energy X-ray Spectrometer (SoLEXS) experiment \citep{sankar2011} with very high spectral resolution would be quite useful to understand the small-scale pre-flare activity. 

\acknowledgments
We acknowledge RHESSI, TRACE, SOHO, and GOES for their open data policy. RHESSI and TRACE are NASA's small explorer missions. SOHO is a joint project of international cooperation between the ESA and NASA. A.V. acknowledges the European Community Framework Programme 7, High 
Energy Solar Physics Data
in Europe (HESPE), grant agreement No.: 263086, and the Austrian Science 
Fund (FWF): P24092-N16. We thank Jeongwoo Lee (NJIT, Newark, USA), Nandita Srivastava (USO/PRL, Udaipur, India), and T. Magara (Kyung Hee University, Yongin, South Korea) for useful discussions. We sincerely thank the anonymous referee for providing very constructive comments and suggestions which
have enhanced the quality and presentation of this paper.


\end{document}